\documentclass[english]{llncs}

\usepackage{amssymb} 
\usepackage{multirow}
\usepackage{float}
\usepackage{graphics}
\usepackage{float}
\usepackage{graphicx}
\usepackage[dvipsnames]{xcolor}
\usepackage{array}
\usepackage{booktabs}
\usepackage{cite}
\usepackage{color}
\usepackage{hyperref} 
\usepackage{cleveref}
\usepackage{rotating}

\newcolumntype{P}[1]{>{\centering\arraybackslash}p{#1}}

\newcolumntype{M}[1]{>{\centering\arraybackslash}m{#1}}
\title{The Worrisome Impact of an Inter-rater Bias on Neural Network Training}

\author{Or Shwartzman\inst{1,2} \and Harel Gazit\inst{1,2} \and Ilan Shelef\inst{2,3,4} \and Tammy Riklin-Raviv\inst{1,2}}
\institute{Electrical and Computer Engineering Department, Ben-Gurion University \and The Zlotowski Center for Neuroscience, Ben-Gurion University \and Department of Radiology, Soroka Medical Center \and Department of Health Sciences, Ben-Gurion University}

\begin{document}

\maketitle

	\begin{abstract}

The problem of inter-rater variability is often discussed in
the context of manual labeling of medical
images.  It is assumed to be bypassed by automatic {\bf model-based} image segmentation approaches,  which are considered `objective',
providing single, deterministic solutions. 
However, the emergence of data-driven approaches such as Deep
Neural Networks (DNNs) and their application to supervised semantic
segmentation - brought this issue of raters' disagreement back to the
front-stage. 

 In this paper, we highlight the issue of inter-rater bias as opposed to random inter-observer variability and demonstrate its influence on
DNN training, leading to
different segmentation results
for the same input images. 
In fact,  lower overlap scores  (e.g. DICE scores) are obtained between the outputs of a DNN  trained on annotations of one rater and tested on another. 
Moreover, we demonstrate that inter-rater bias in the training examples is amplified and become more consistent when
considering the segmentation predictions of the DNNs' test data. 
We support
our findings by showing that a classifier-DNN trained to distinguish
between raters based on their manual annotations performs better 
when the
automatic segmentation predictions rather than the actual raters' annotations were tested.\\

For this study, we used two different datasets: the ISBI 2015
Multiple Sclerosis (MS) challenge dataset, which includes MRI scans each with annotations provided by two 
raters with different levels of
expertise~\cite{carass2017longitudinal}; and Intracerebral Hemorrhage (ICH) CT scans with manual and semi-manual segmentations~\cite{hershkovich2016probabilistic}. 
The results obtained allow us to underline a worrisome clinical
implication of a {\em DNN bias induced by an inter-rater bias} during training. 
%
%
%
Specifically, we present a consistent underestimate of MS-lesion
loads when calculated from segmentation predictions of a DNN trained
on input provided  by the less experienced rater. In the same manner, the differences in ICH volumes calculated based on outputs of identical DNNs , each trained on annotations from a different source 
are more consistent and larger than the differences in volumes between the manual and semi-manual annotations used for training.

	\end{abstract}


	\section{Introduction \label{sec:Introduction}}

	Semantic image segmentation plays an essential role in
        biomedical imaging analysis. It is considered a challenging
        task not necessarily due to moderate image quality but mainly
        since it is not well defined. Different human raters and even
        the same rater at different time points may draw the boundary
        of a particular region of interest (ROI) in different
        manners, see e.g.,~\cite{Hayward2008, Joskowicz2019}. 
To address this well-known issue, an Expectation Maximization (EM)
framework has been suggested for simultaneous evaluation of the
raters' performance level and the consensus
segmentation~\cite{warfield2004simultaneous}. 
In~\cite{gordon2009evaluation} the variability of experts annotations was
discussed in the context of segmentation evaluation. 
Standardizing measures to assess
human observer variability for clinical studies was suggested in~\cite{Popovic2017}. 

Machine vision algorithms are often praised for being repeatable and objective. This is, however, true only when considering deterministic, model-based approaches and obviously, the resulting segmentation is model-dependent.  
	The emergence of machine learning and deep learning, in particular, have made data-driven approaches dominant. A supervised deep
        neural network (DNN) for image segmentation is trained by
        input images and their corresponding manual annotations that are used
        for calculating the network's loss. Backpropagation guided by
        the loss enables implicit modeling, learned
        indirectly from the conditional distribution of the data. This
        process has
        shown to be very powerful, outperforming model-based
        segmentation approaches, as it allows us to generalize from seen to unseen data. Nevertheless,
        in most benchmark datasets, while the images to segment
        vary, the annotation is often done by a single
        annotator. Therefore, it is not unlikely that it is the
        annotator's subjective outlook or bias that actually shapes the
        segmentation model and consequently influences the resulting image
        labels.

	While the issue of inter-rater bias has significant implications,
        in particular nowadays, when an increasing number of deep
        learning systems are utilized for the analysis of clinical
        data~\cite{Litjens17,Shen17}, it often seems to
        be neglected. In~\cite{Jungo2018} deep learning is used
        to analyze the effect of common
       label fusion techniques on the estimate of segmentation
       uncertainty among observers. Nevertheless, the problem
       addressed there is completely different and the distinction
       between random versus consistent (i.e., bias) inter-observer variability is not made. 
Consistent differences in medical imaging data have been addressed for
a related problem of inter-site variability 
        by~\cite{styner2002multisite} for structural MRI and
        by~\cite{mirzaalian2015harmonizing} for diffusion MRI. 
          In addition, inter-site variability was also investigated in
          the context of deep-learning in~\cite{gibson2018inter}. Yet,
          inter-site variability refers mainly to the differences in
          imaging data acquired by different machines and not the raters.

	The main objective of this work is to quantify the impact of
        the observer's conception and competence on segmentation
        predictions by a supervised deep learning framework. In a sequence of
        experiments, utilizing both classifier- and segmenter-DNNs,
        we show that inter-rater bias affects significantly the DNN
        training process, and therefore leads to different
        segmentation results for the same test data, despite carefully
        using an identical DNN architecture and training regime. 
In fact, consistently lower Dice scores are calculated if training and
test segmentations are of different raters. 
We support
our findings by training a classifier DNN to distinguish between different
raters based on their segmentations. Surprisingly, much more significant
rater-classification results were obtained when the segmentation
predictions (the outputs of the segmentation DNNs) rather than
the manual annotations (used for training) were considered. 
We then suggest a compromise training regime, that incorporates
the segmentations of both raters.

For the purpose of this study, we used two different datasets. This includes a multi-modal MRI dataset of Multiple Sclerosis (MS)  patients, that was made publicly available by the ISBI 2015 MS-lesion challenge~\cite{carass2017longitudinal} and a CT dataset of Intracerebral Hemorrhage (ICH) patients from a private source~\cite{hershkovich2016probabilistic}. 
Each scan in both datasets is provided with annotations from two sources. The MS-lesion scans were annotated by two raters with different levels of expertise and the  ICH scans were annotated manually and in a semi-automatic (semi-manual) manner with an interactive segmentation tool~\cite{hershkovich2016probabilistic}. The interactive tool is based on proposal segmentation generated automatically, following its correction based  on mouse clicks provided by the user in 
The results obtained allow us to highlight a worrisome clinical
implication of DNN bias induced by inter-rater bias during training. 
Specifically, relative underestimation of the MS-lesion load by the less experienced rater
was amplified and became consistent when the volume calculations were based on the segmentation predictions of the DNN that was trained on this rater's input. 
In the same manner, the differences in ICH volumes calculated based on outputs of identical DNNs, each trained on annotations from a different source 
were more consistent and larger than the differences between the ICH volumes as calculated based on the manual and semi-manual annotations used for training.

The rest of the paper is organized as follows.
In Section~\ref{sec:Methods} we describe the data; the segmenter and
classifier DNNs as well as the evaluation measures we
used. In Section~\ref{sec:Experiments} we present the experimental results. We conclude in Section~\ref{sec:Summary}.

	\begin{figure}[t]
		\centering
			\begin{tabular}{ccc}
				\includegraphics[width=3.7cm]{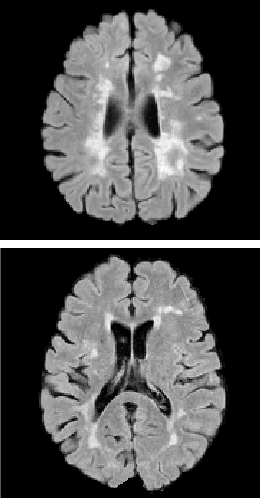}
                          & \includegraphics[width=3.7cm]{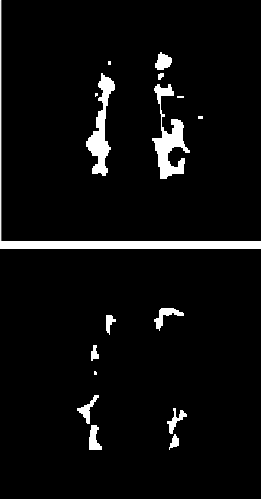}
                          & \includegraphics[width=3.7cm]{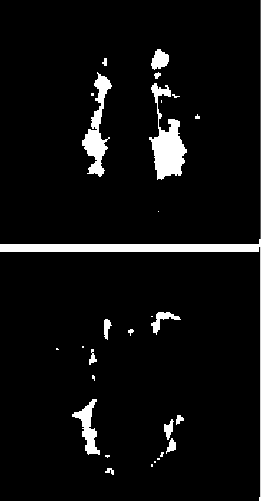}
                          \\
	(a) Brain scans & (b) source~\#1 & (c) source~\#2 
			\end{tabular}
\caption{MRI brain scans of MS patients annotated by two different raters. (a) 2D slices of 3D  brain scans (FLAIR) (b-c) MS annotations of rater~\#1 and rater
                 ~\#2, respectively. \label{figure:MS}}

	\end{figure}

\begin{figure}[t]
	\centering
	\begin{tabular}{ccccc}
		\includegraphics[width=3.5cm]{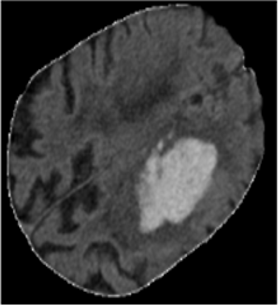}
		& \includegraphics[width=3.5cm]{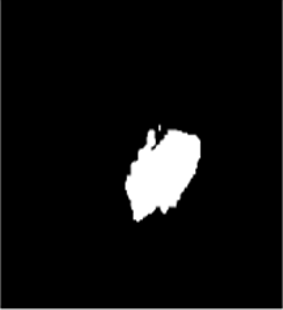}
		& \includegraphics[width=3.5cm]{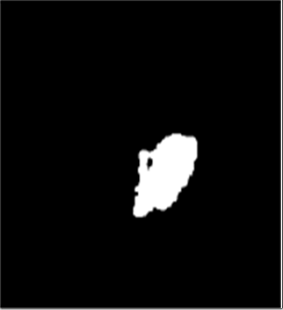}
		\\
		(a) Brain scans & (b)source~\#1 & (c) source~\#2
	\end{tabular}
	\caption{A CT brain scan of an ICH patients with two different
		annotations. (a) A 2D slices of a 3D CT brain scan
		(b-c) ICH segmentations obtained by manual (source~\#1) and semi-manual (source~\#2) annotations.  \label{figure:CT}}

\end{figure}

	\section{Methods}
\label{sec:Methods}

	\subsection{Data}
	\subsubsection{Multiple sclerosis (MS) Lesion MRI scans:}
	The multi-modal brain MRI dataset consists of 21 scans of MS lesion patients 
 from the ISBI 2015 MS segmentation
challenge dataset~\cite{carass2017longitudinal}. Scans were acquired by a 3T Philips scanner
and include T1-w, T2-w, PD-w and FLAIR sequences.  
MS-lesions were annotated by two different raters with four
        (source~\#1) and ten (source~\#2) years of experience in delineating
        lesions. The second rater has overall 17 years of experience
        in structural MRI analysis. More details can be found in~\cite{carass2017longitudinal}.
Fig.~\ref{figure:MS} presents two MRI slices along with the labels of
both raters from the MS-lesion dataset. 
	\subsubsection{Intracerebral Hemorrhage (ICH) CT scans:}
The brain CT dataset includes scans of 	28 ICH patients 
 acquired at the Soroka Medical Center with Philips Brilliance CT 64 system without radiocontrast agents injection. The size of each scans is  $512 \times 512 \times [90-100]$ voxels with voxel size of $0.4\times 0.48 \times 3$  mm$^3,$ with 1.5 mm overlap in the axial direction. Each scan was annotated by an experienced radiologist  - manually (source~\#1) and with the help of a semi-manual segmentation tool (source~\#2) \cite{hershkovich2016probabilistic}. Fig.~\ref{figure:CT} presents a CT slice from the ICH dataset along with the manual and the semi-manual segmentations. 

\subsubsection{Data patches:} Due to GPU memory limitations, we partitioned the data into overlapping patches.  The ICH dataset was partitioned into cubic patches of size $100\times100\times 100$ voxels and the multi-modal MS-lesion data was partitioned to 4D patches of size $100\times100\times 100\times 4$ voxels.

	\subsection{Neural Networks: Architectures and Training}
\begin{figure}[h!]
	\includegraphics[width=15.3cm]{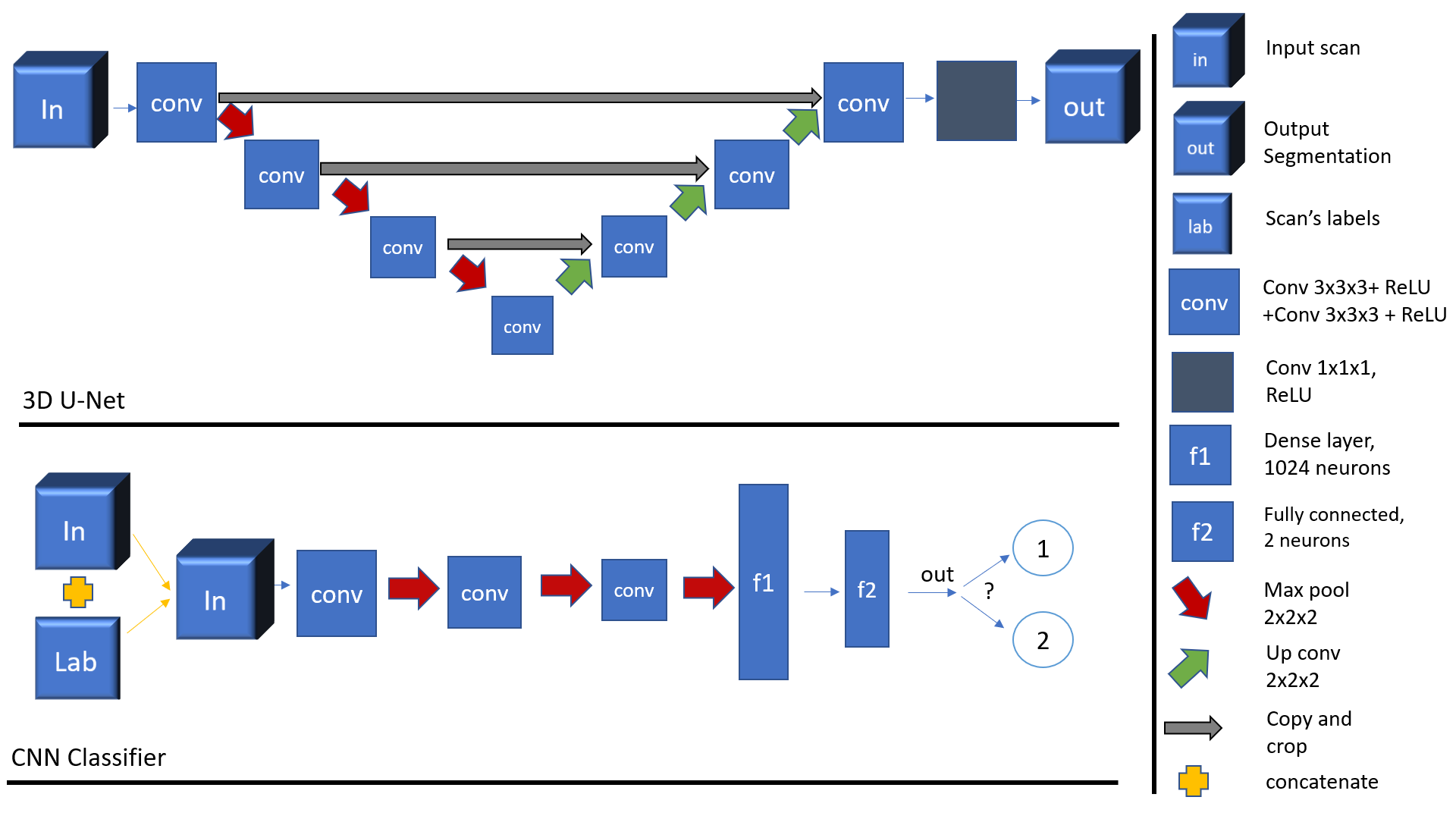}
\caption{DNNs architectures: upper panel: 3D U-Net for MS-lesion and brain ICH segmentation; lower panel: Classifier DNN for classifying the source of the segmentations \label{fig:Architectures}}

	\end{figure}

	The architectures of both the 3D U-Net - used for segmentation and the
        classifier CNN - used for rater identification, are illustrated in Figure~\ref{fig:Architectures}.\\
	{\bf 3D U-Net.}
The U-Net is a symmetrical fully convolutional DNN with skip
connections between its down-sampling and up-sampling paths~\cite{ronneberger2015u}.
	For this study, we implemented a 3D U-Net \cite{cciccek20163d}
with Tensorflow and trained it for the MS-lesion segmentation of multi-modal 3D brain MRI scans (dataset \#1), and for the ICH segmentation of 3D brain CT scans (dataset \#2).
	For the MS-lesion dataset, we used a cross-validation scheme with 11, 5 and 5 brain scans for training, validation, and testing, respectively, while for the ICH dataset we used a cross-validation scheme with 16, 4 and 8 brain scans for training, validation and testing, respectively.
	In both cases, training was based on the cross-entropy loss. As we mentioned before, due to GPU memory limitations, we used patches. As in the original U-Net
        paper~\cite{ronneberger2015u}, we favored large input patches over a large
        batch size and hence set the batch size to two with a learning rate of
        0.0005. We also used batch
normalization and dropout of 0.5. \\
	{\bf CNN Classifier.}
	The CNN classifier is used for rater identification based on either the actual annotations provided by the raters or based on the automatic segmentations provided by the U-Nets, where each network was trained on the annotations provided by either of the raters.  The classifier's architecture consists of 4 consecutive blocks, followed by a dense layer and a fully connected layer. Each block is comprised of 2 convolutions followed by one max pooling (besides the last one).
We trained and tested the CNN classifiers based on volumetric ($100 \times 100 \times 100$) patches extracted from the MS-lesion (21 brain scans) and the ICH (28 brain scans) datasets. 
Specifically, for the MS-lesion dataset, we used 7000 patches (11 brain scans) for training, 2800 patches for validation, and 2500 patches (5 brain scans) for the test phase. For the ICH dataset, we used 6914 patches (16 brain scans) for training, 2500 patches for validation, and 2600 patches (6 brain scans) for the test phase. 
As the sizes of the data sets were relatively small we
ran these experiments with 2-fold cross validation, switching between the training, the test,
and the validation sets. In addition, we ran each experiment multiple times. 
	We used the cross entropy loss, a batch size of only
        1, dropout of 0.5, batch normalization and set the learning rate to $0.001$. 
	
	\subsection{Quantitative evaluation measure}
	We use the Dice score \cite{dice1945measures} to
        quantify the compatibility between the manual segmentations of
        the two raters and to compare between the raters'
        annotations and the segmentations generated by two
        3D U-Nets, where each was exclusively trained on the manual segmentations of either of them. 
        We also present the accuracy of the classifier-CNN defined as
        the number of correctly classified segmentation patches per-brain (source~\#1 or
        source~\#2) with respect to the total number of patches.   
	
	\section{Experiments}
	\label{sec:Experiments}
To assure that the results obtained are not affected by the random initialization of the network's weights, we trained the U-Net for each dataset and annotation source several times. Segmentation and classification results presented in the paper were obtained by averaging the DNNs' outputs.  
	\subsection{MS-lesion and Intrcerebral Hemorrhage (ICH) segmentation for cross-evaluation}	
     The mean Dice score between the manual
        segmentations of source~\#1 and the corresponding manual segmentations of source~\#2 is $0.7341\pm 0.0967$ for the MS-lesion dataset, and $0.83\pm 0.05$ for the ICH dataset, indicating significant mismatches between the sources. These mismatches can be visually observed in Figs.~\ref{figure:MS}-\ref{figure:CT}. To demonstrate the impact of the
        different raters' segmentations on the DNN training, we trained
        a 3D U-Net twice, each time with the segmentations of either
        of the raters. For the sake of convenience, we term the U-Nets
        trained on the segmentations of source~\#1 and source~\#2 by network~\#1 and network~\#2, respectively, although the very same
        architecture and training regime were used. We then tested the U-Net performances by
        comparing the manual segmentations of each of the raters with
        the output segmentations obtained for each of the training
        sessions. Thus, for each dataset, we performed four comparisons, either training and testing using the same source segmentations or training on one source and testing on the other (cross rater evaluation). \\
 {\bf Results:}\\
Figure~\ref{tbl:table_of_figures} presents
 slices of 3D MRI scans from the MS lesion dataset (rows 1-2) and of CT scans from the ICH dataset (rows 3-4) along with the segmentations provided by two different sources (source~\#1 and source~\#2, columns 1-2) as well as the segmentations provided by DNNs (network \#1 and network~\#2) each is trained on segmentations from a different source (columns 3-4). 
 Note the
        relative visual similarity between the manual segmentations of source
       ~\#1~(\#2) and the DNN segmentations of network~\#1~(\#2). 
Cross-evaluation Dice scores obtained by training based on source~\#1 and testing
based on source~\#2 and vice versa are presented in rows 1-2 of
Table~\ref{tbl:CE} and Table~\ref{tbl:CE_CT} for the MRI and CT datasets, respectively. For comparison, the Dice
scores obtained for training and testing with segmentations provided by the 
same rater are presented as well. 
Two-sample t-tests with p-values of $9.9029e-04$ for the MRI and $0.0167$ for the CT datasets, indicate 
statistically significant differences between
the Dice scores obtained for same-source experiments (in which segmentations of the same rater were used for both training and test) and cross-source experiments (in which the networks trained on the segmentations of one rater were tested on the segmentations of the other rater).\\
\subsection{Mixed Source Training}
	\textbf{Experimental design.}
	In this experiment, we aimed to simulate a situation in which the number of training annotations provided by an expert is much smaller than the number of annotations provided by a less experienced annotator.
	We, therefore, trained a 3D U-Net multiple times with a mixed training set, containing a different number of annotations of the two raters.
	 We then used the test sets to check the dependency between the Dice scores, calculated with respect to annotations provided by one of the sources,  and the number of training annotations of that source. \\
	\textbf{Results:}\\
For both the ICH and the MS datasets, we used the entire datasets provided by source~\#1 and only part of the annotations provided by source~\#2. 
	 The results, for the MS dataset, are presented in rows 3-5 of Table~\ref{tbl:CE}. 
The main finding is that the average Dice score, with respect to source~\#2 decreases from 0.8 	(when the U-Net is exclusively trained on input from source~\#2) to 0.78  when the U-Net is trained on input from source~\#1 and only one third of the input from source~\#2.
Similar findings are obtained for the ICH dataset, as presented in rows 3-6 of Table~\ref{tbl:CE_CT}. 
Here, the average Dice score, with respect to source~\#2 decreases from 0.86 	(when the U-Net is exclusively trained on input from source~\#2) to 0.84 when the U-Net is trained on input from source~\#1 and less than one third of the input from source~\#2.


\subsection{MS-lesion load and ICH volume}
\begin{figure}[t!]
	\includegraphics[width=\textwidth]{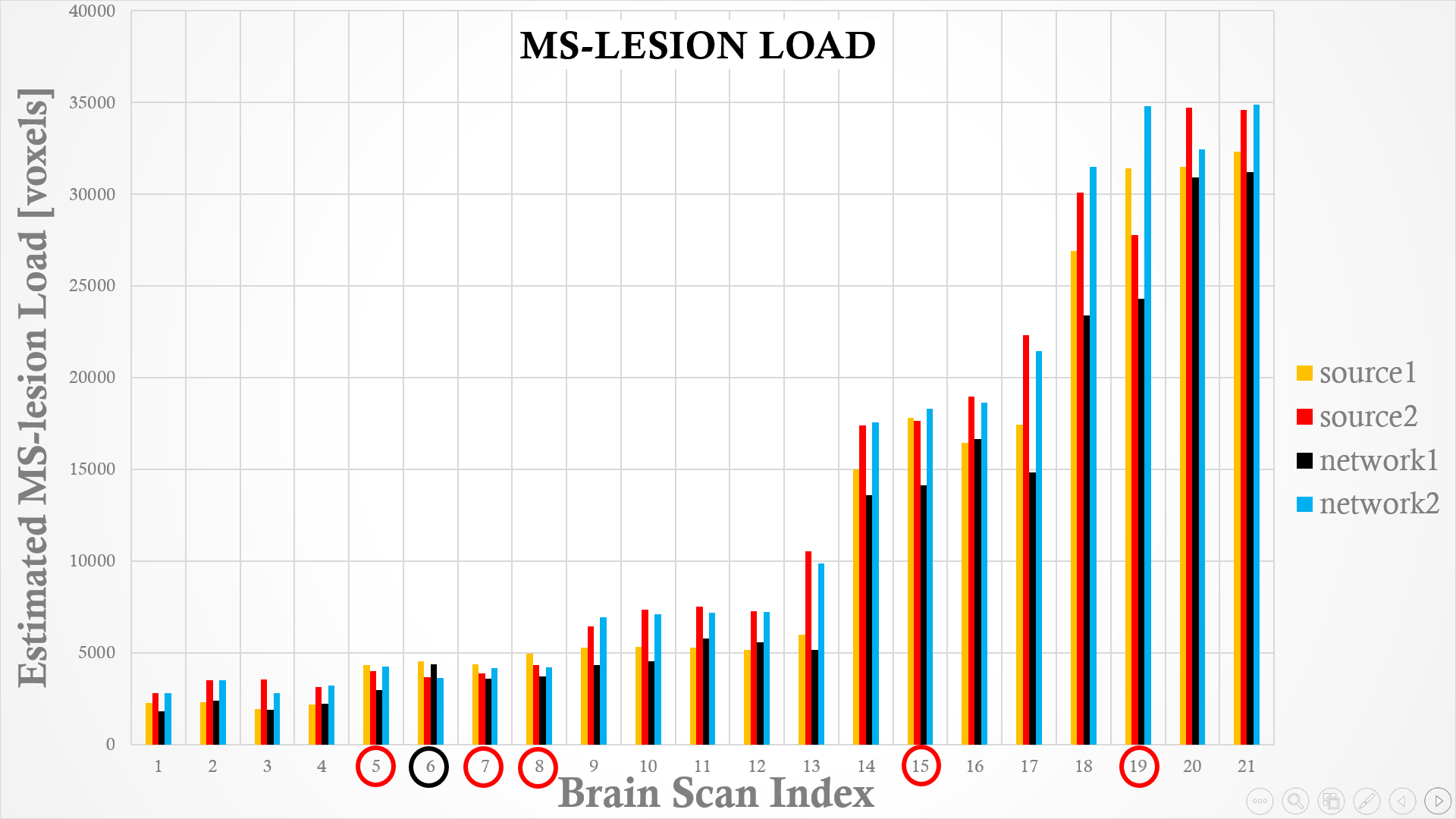}
\caption{Comparison of MS-lesion load estimate}
	\label{fig:ms-load}

\end{figure}

An important clinical measure is the MS lesion load which indicates the severity of the disease and may predict
long-term cognitive dysfunction in MS patients~\cite{Patti2015}. The
lesion load is estimated by the sum of voxels labeled as lesion.
Figure~\ref{fig:ms-load} presents a bar plot of the lesion loads of 21 brain MR volumes as calculated from the segmentations of the raters and the networks. For clarity, brain indices (x-axis) were arranged in an increasing order of the estimated MS-lesion load. The plot shows that the lesion loads calculated from the manual segmentations from source~\#1 (yellow) are, for most brains (all of them but 6, circled in red/black), lower than the lesion loads calculated from the manual segmentations provided by source~\#2 (red). However, the alerting results refer to the network segmentations. MS-load estimations based on network~\#1 (black) segmentations are almost consistently (all of the brains but 1, circled in black) much smaller than the load estimations of network~\#2 (light blue), showing that inter-rater bias induces an increased and generalized bias between the networks.

\begin{figure}[t!]
	\includegraphics[width=\textwidth]{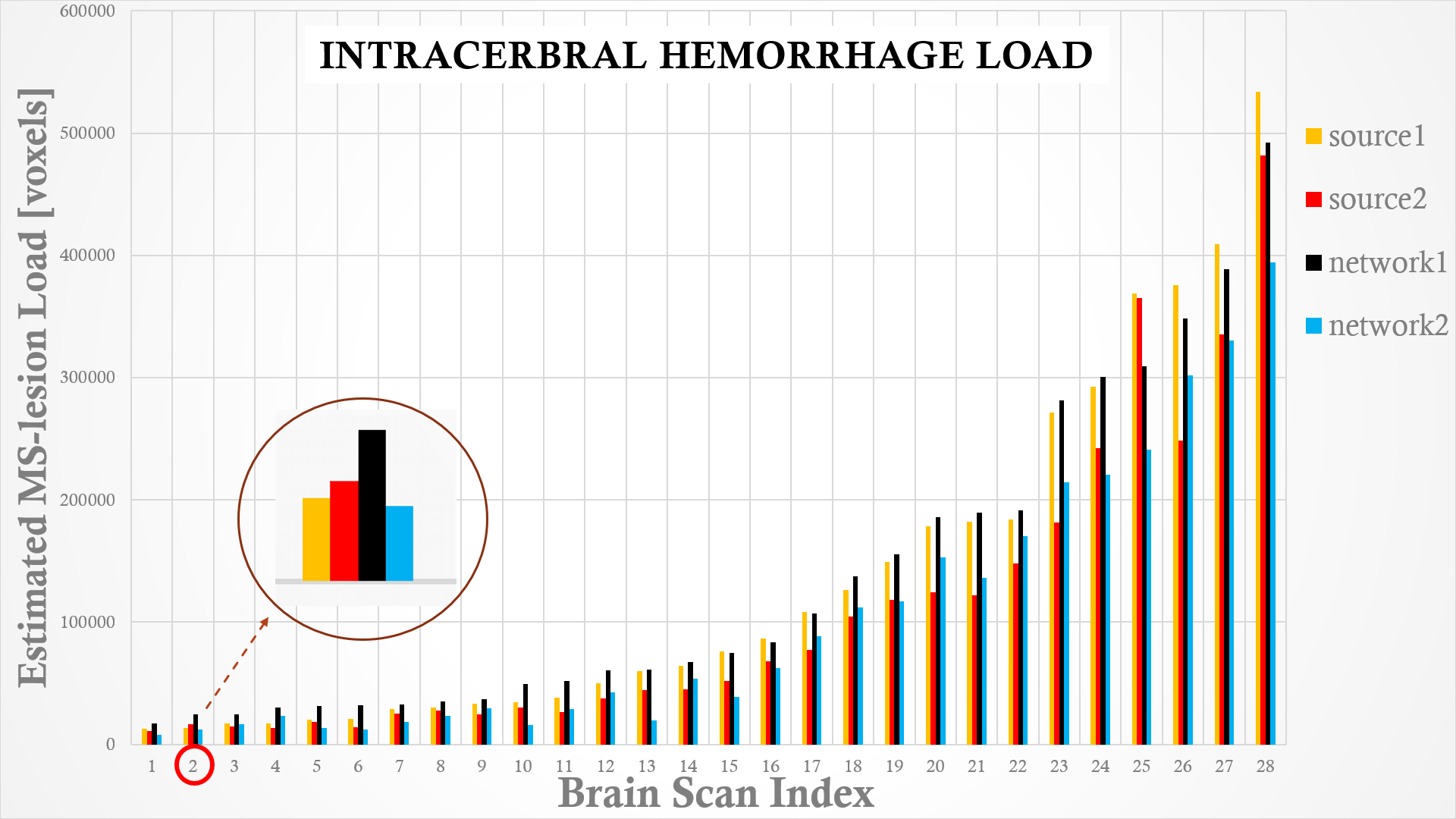}
	\caption{Comparison of ICH load estimate}
	\label{fig:intra-load}

\end{figure}

Another important clinical measure is the ICH volume, which is a key factor experts take under consideration for deciding whether a patient needs to undergo a surgery or not. The load is estimated by the sum of voxels labeled as lesion. Figure~\ref{fig:intra-load} presents a bar plot of hemorrhage volume of 28 brain CT volumes as calculated from the segmentations of the raters and the networks. The plot shows that the ICH volumes calculated from the  segmentations from source~\#1 (yellow) are, for most brains (all of them but 1, circled in red), higher than the ICH volumes calculated from source~\#2 (red). The figure also shows that when calculated by the networks, the hemorrhage volume is consistently (28/28) higher according to network~\#1(black), which corresponds with the results obtained in the MS-load experiment.

	\newcommand{\addpicone}{\includegraphics[width=9em]{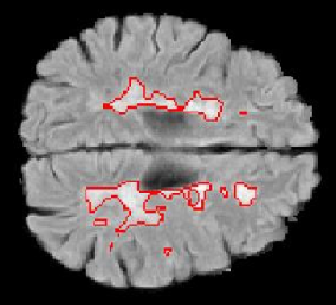}}\quad
	\newcommand{\addpictwo}{\includegraphics[width=9em]{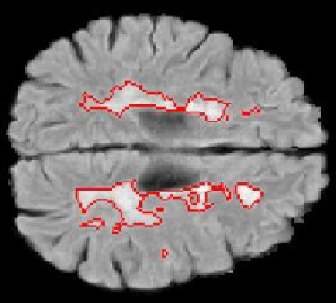}}
	\newcommand{\addpicthree}{\includegraphics[width=9em]{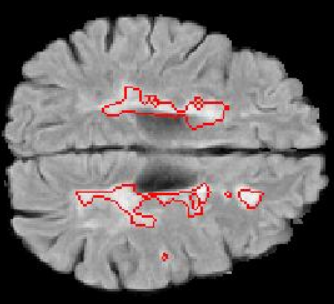}}
	\newcommand{\addpicfour}{\includegraphics[width=9em]{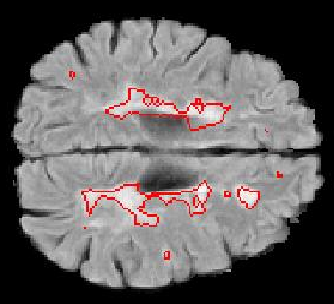}}
	\newcommand{\addpicfive}{\includegraphics[width=9em]{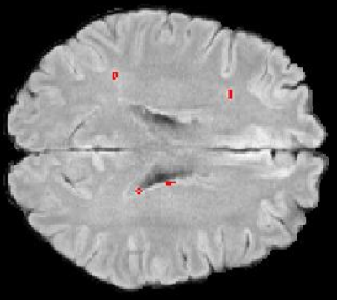}}\quad
	\newcommand{\addpicsix}{\includegraphics[width=9em]{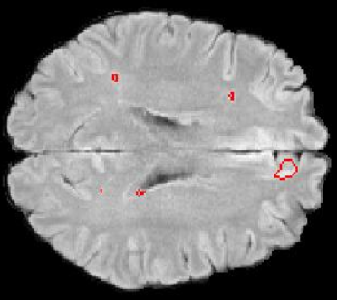}}
	\newcommand{\addpicseven}{\includegraphics[width=9em]{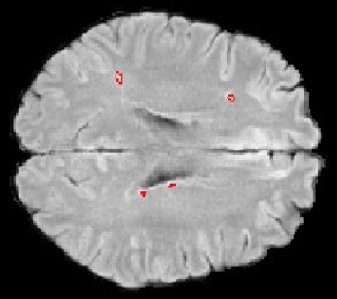}}
	\newcommand{\addpiceight}{\includegraphics[width=9em]{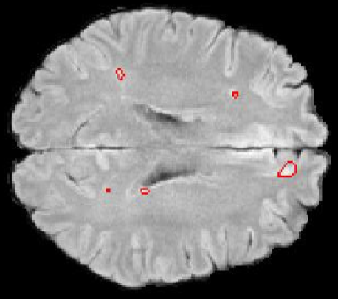}}
	\newcommand{\addpicnine}{\includegraphics[width=9em]{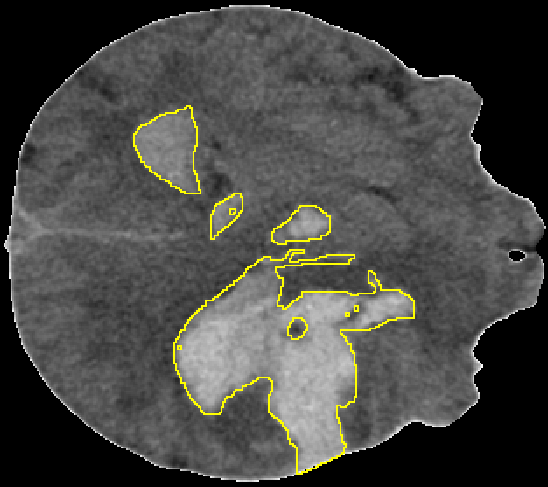}}\quad
	\newcommand{\addpicten}{\includegraphics[width=9em]{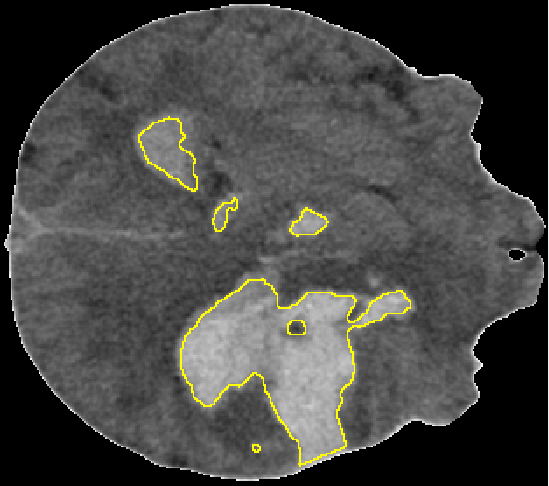}}
	\newcommand{\addpiceleven}{\includegraphics[width=9em]{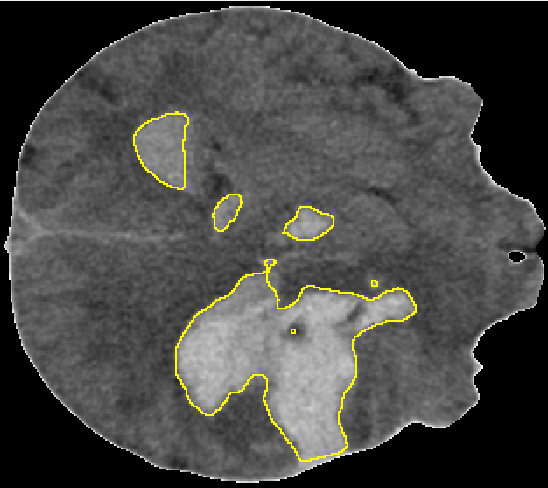}}
	\newcommand{\addpictwelve}{\includegraphics[width=9em]{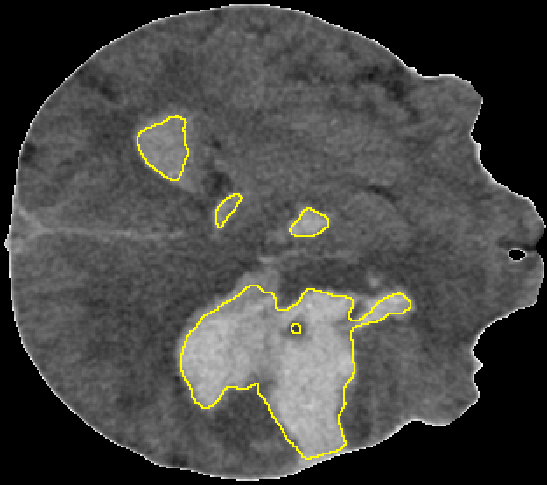}}
	\newcommand{\addpicthirth}{\includegraphics[width=9em]{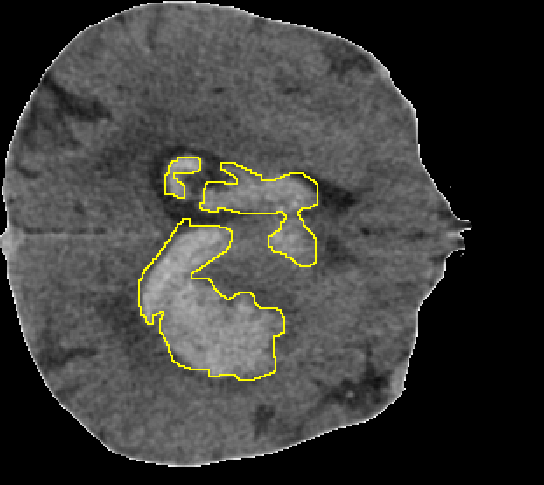}}\quad
	\newcommand{\addpicfourth}{\includegraphics[width=9em]{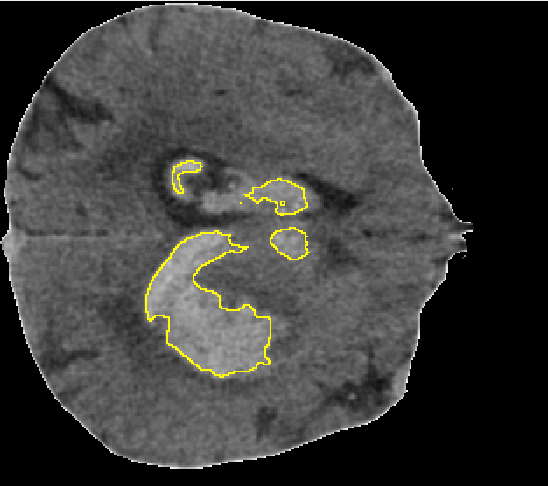}}
	\newcommand{\addpicfifth}{\includegraphics[width=9em]{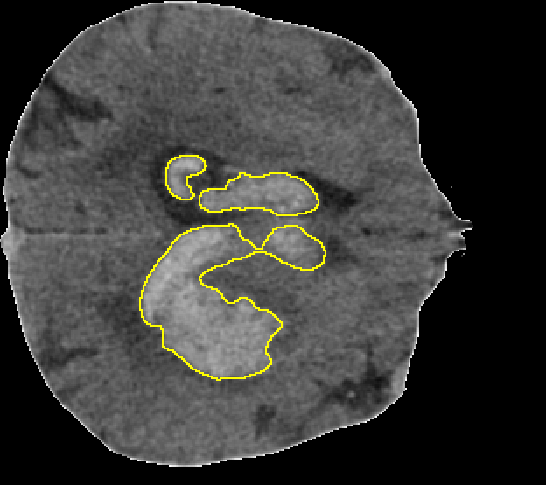}}
	\newcommand{\addpicsixt}{\includegraphics[width=9em]{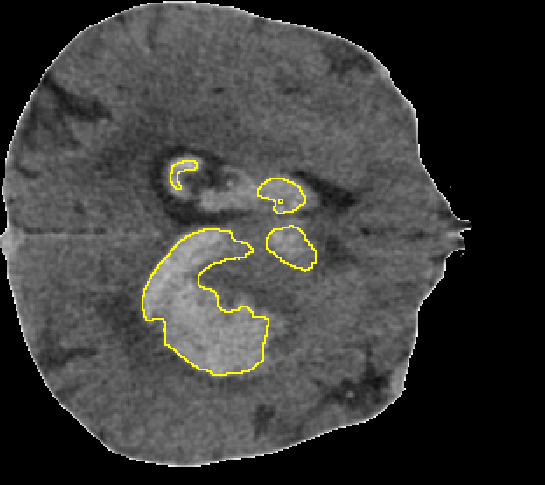}}

	\begin{figure}[h!]
		\centering
		\begin{tabular}{lcccc}
			\begin{turn}{90} ~~MRI, MS-lesion \end{turn} & \addpicone & \addpictwo & \addpicthree & \addpicfour \\
			\begin{turn}{90} ~~MRI, MS-lesion \end{turn} & \addpicfive & \addpicsix & \addpicseven & \addpiceight \\
			\begin{turn}{90} ~~~~~~CT, ICH \end{turn}  & \addpicnine & \addpicten & \addpiceleven & \addpictwelve \\
			\begin{turn}{90} ~~~~~~CT, ICH \end{turn}  & \addpicthirth & \addpicfourth & \addpicfifth & \addpicsixt \\
	&(a) source~\#1&(b) source~\#2&(c) network~\#1&(d) network~\#2
		\end{tabular}
\caption{MRI and CT scans of MS-Lesion and ICH patients, respectively, along with the manual (or semi-manual)
  and the U-Net segmentation contours. (a) Segmentations provided by
  source~\#1 (b) Segmentations provided by
  source~~\#2 (c) Automatic
  segmentations of the U-Net, trained with the data labeled by the
  source~\#1 (d) Automatic segmentations of the U-Net trained with the
  data labeled by source~\#2. \label{tbl:table_of_figures}}

	\end{figure}

	\begin{table}[b!]
		\centering
		\caption{MRI dataset: Dice scores of cross evaluation;
                  same-rater and multi-rater}
		\label{tbl:CE} 
		\begin{tabular}{P{3.5cm}  P{3.7cm}  P{3.7cm} } 
			\centering 
			Trained/Evaluated & rater 1 & rater 2 \\ 
			\midrule 
			rater 1 & $0.82\pm0.05$ & $0.74\pm0.08$ \\ 
			rater 2 & $0.72\pm0.10$ & $0.82\pm0.07$ \\
	        raters2+1 (ratio 11/11) & $0.8\pm0.08$ & $0.79\pm0.09$\\
	        raters2+1 (ratio 05/11) & $0.79\pm0.06$ & $0.77\pm0.06$\\
	        raters2+1 (ratio 04/11) & $0.78\pm0.05$ & $0.77\pm0.04$ 
		\end{tabular}
	\end{table}

\begin{table}[b!]
	\centering
	\caption{CT dataset: Dice scores of cross evaluation;
		same-rater and multi-rater}
	\label{tbl:CE_CT} 
	\begin{tabular}{P{3.5cm}  P{3.7cm}  P{3.7cm} } 
		\centering 
		Trained/Evaluated & source 
		\#1 & source~\# 2 \\ 
		\midrule 
		source~\# 1 & $0.86\pm0.1$ & $0.79\pm0.07$ \\ 
	source~\# 2 & $0.80\pm0.09$ & $0.87\pm0.07$ \\
		sources 1+2 (ratio 16/16) & $0.86\pm0.07$ & $0.82\pm0.05$\\
		sources 1+2 (ratio 8/16) & $0.84\pm0.07$ & $0.81\pm0.06$\\
		sources 1+2 (ratio 06/16) & $0.83\pm0.06$ & $0.81\pm0.06$\\
	sources 1+2 (ratio 05/16) & $0.82\pm0.10$ & $0.80\pm0.07$
	\end{tabular}

\end{table}

\subsection{Rater and network classification}
	\label{sec:Exp_class}
	\textbf{Experimental design.}
To support our finding, we trained classifier DNNs to differentiate
between the  different segmentation sources based on the actual annotations. The
input to each of the classifiers include pairs of images (CT or MRI) and the corresponding
segmentations, randomly selected from either of the sources. The classifier
was also used to distinguish between segmentation predictions of
network~\#1 and network~\#2 which were trained based on the
segmentations of source~\#1 and source~\#2, respectively. \\
\textbf{Results:}\\
The first row in Table~\ref{tbl:class} (MS-lesion data) and in Table~\ref{tbl:class_CT} (ICH data) presents source classification results
based on the respective manual annotations. The successful classifications 
(hit rates of 0.87 and 0.917, respectively) indicate consistent differences between the segmentations provided by the different sources.  
The second row in each of the tables presents much better classification
results (hit rates of 0.9 and 0.95, respectively) when the classifier CNNs are tested on
segmentation predictions provided by the U-Nets (network~\#1 and network~\#2),
rather than the source segmentations themselves.  This implies that not only do the segmentations provided by each source have particular characteristics that can be distinguished by a classifier CNN, but also these distinguishing characteristics seem to be enhanced (or become more consistent) in the automatic segmentations of network~\#1 and network~\#2.

\begin{table}[t!]
\caption{{\bf Classification results for MS dataset.}  Mean classification hit-rates ($\pm$ std) obtained by the classifier CNN for distinguishing between  segmentations provided by the two sources (first row) and by the two U-Nets, where each was trained on either of the sources (second row).} 
		\label{tbl:class}
		\centering			
                \begin{tabular}{lcc}
	& ~~\small{\# of patches} &~~\small{classification hit-rate $\pm$ std} \\
\hline
source \#1 vs. \#2 & 8441 & $0.90\pm0.05$ \\
network \#1 vs. \#2 & 8597 & $0.93\pm0.03$ 
\end{tabular}

\end{table}

\begin{table}[t!]
	\caption{{\bf Classification results for the ICH data.}  each classification hit-rates ($\pm$ std) obtained by the classifier CNN for distinguishing between  segmentations provided by the two sources (first row) and by the two U-Nets, where each was trained on either of the sources (second row).}
	\label{tbl:class_CT}
	\centering			
	\begin{tabular}{lcc}
		& ~~\small{\# of patches} &~~\small{classification hit-rate$\pm$ std} \\
		\hline
		 source \#1 vs. \#2 & 9260 & $0.917\pm0.06$ \\
		 network \#1 vs. \#2 & 9494 & $0.95\pm0.04$ 
	\end{tabular}

\end{table}

	\section{Conclusions and Discussion}

	\label{sec:Summary}
In this paper, we highlighted the problem of inter-rater bias in medical image segmentation, which is often overlooked in the context of deep learning methods. Specifically, we exemplified the phenomenon of training-induced bias using CT and MRI datasets of ICH and MS-lesion patients scans (respectively) where each of these datasets was annotated by two different sources. While one could expect 
that DNNs trained with different target segmentations would converge in a different manner thus providing different test results, the amplification of the differences between the DNNs' outputs was surprising. These findings are worrisome. MS-lesion loads are used for evaluating MS disease progress. Patient's ICH volume estimates are critical to the determination of the therapy procedure, which may
involve surgery in addition to medicine intake. 
While the results shown are dataset-specific 
we believe that the phenomena of {\it DNN induced bias} is a general
one, and as such has clinical
implications that the biomedical imaging community should be aware of. Nowadays, when
much effort is made to improve DNNs performances,
the clinical training of the human annotator or the accuracy of the annotating source, that provided the `ground truth' segmentations
to the network, should be also considered.  

Our study demonstrates that the expertise of the annotator directly influences DNN's training and consequently its test results. Since expert's annotations are costly and often only partially available we suggest a mixed training process, using annotations provided by two or more sources. Addressing the common situation in which a less experienced rater provides most of the annotated data, we show that it is sufficient to use a small portion of expert's annotations during training to influence DNN's performances.

\bibliographystyle{splncs04}

\bibliography{inter_rater}

\end{document}